\documentclass[12pt]{article}
\usepackage[top = 2 cm, bottom = 3 cm, left = 3 cm, right = 2 cm]{geometry}
\usepackage{authblk, cite, color, amssymb, amsmath}
\usepackage[colorlinks = true, linkcolor = blue, citecolor = red]{hyperref}
\usepackage[dvipsnames]{xcolor}

\begin{document}
\title{\bf Octonionic geometry and conformal transformations}
\author{Merab Gogberashvili}
\affil{\small Javakhishvili Tbilisi State University, 3 Chavchavadze Avenue, Tbilisi 0179, Georgia \\
Andronikashvili Institute of Physics, 6 Tamarashvili Street, Tbilisi 0177, Georgia \\
E-mail: {\it gogber@gmail.com}}
\maketitle

\begin{abstract}
We describe space-time using split octonions over the reals and use their group of automorphisms, the non-compact form of Cartan's exceptional Lie group G2, as the main geometrical group of the model. Connections of the G2-rotations of octonionic 8D space with the conformal transformations in 4D Minkowski space-time are studied. It is shown that the dimensional constant needed in these analysis naturally gives the observed value of the cosmological constant.
\vskip 1mm

\end{abstract}
\vskip 5mm


The algebra of octonions \cite{Sc, Sp-Ve, Baez} is interesting mathematical object for physical applications (see reviews \cite{Rev-1, Rev-2, Rev-3, Rev-4}). In our previous papers \cite{Gog, Go-Sa, Gog-PP} it was suggested to use split octonions as a universal mathematical structure in physics (instead of vectors, tensors, spinors, etc.), since many properties of the physical world find proper descriptions in terms of the features of the algebra.

In our approach world-lines (paths) of particles are parameterized by the elements of split octonions over the field of real numbers \cite{Gog, Go-Sa},
\begin{equation} \label{s}
ds = d\omega + J_n d\lambda^n + j_n dx^n + cI dt ~~~~~ (n = 1, 2, 3)
\end{equation}
(a pair of repeated upper and lower indices implies a summation), where $t$ and $x^n$ denote the ordinary space-time coordinates, and $\omega$ and $\lambda^n $ are interpreted as the phase (classical action) and the wavelengths associated with the octonionic signals, respectively. The eight octonionic basis units in (\ref{s}) are represented by one scalar (denoted by $1$), the three vector-like objects $J_n$, the three pseudo vector-like elements $j_n$ and one pseudo scalar-like unit $I$,
\begin{equation}
J_n^2 = - j_n^2 = -I^2 = 1 ~.
\end{equation}
The norm (interval) of (\ref{s}) is assumed to be non-negative,
\begin{equation} \label{sN}
N^2 = dsds^\dag = d\omega^2 - d\lambda^2 + dx^2 - c^2dt^2 \geq 0~.
\end{equation}
The second condition is that the norm of the 7D vector part of (\ref{s}),
\begin{equation} \label{V}
V = J_n d\lambda^n + j_n dx^n + cI dt ~,
\end{equation}
which appears in the dominators of the generalized Lorentz transformations \cite{Go-Sa}, to be time-like,
\begin{equation} \label{|V|}
V^2 = c^2 \left(1 - \frac{v^2}{c^2} + \frac{\dot\lambda^2}{c^2}\right) \geq 0~.
\end{equation}
Here $v^i = dx^i/dt$ and $\dot \lambda^i$ are the 3-velocities and the rates of the changes of wavelengths (generalized forces), respectively. Since the extra 'quantum' term in (\ref{|V|}) is positive, velocities of particles with $\dot\lambda^2 > v^2$ exceed speed of light, i.e. they are virtual. Thus we can introduce the additional classification:
\begin{eqnarray} \label{Y-classification}
|dx| &<& |d\lambda| ~, ~~~~~ (virtual~regime) \nonumber \\
|dx| &\geq& |d \lambda| ~. ~~~~~ (observable~regime)
\end{eqnarray}
The identification $\lambda^n \sim \hbar /p^n$, where $p^n$ is the momentum associated with an octonionic signal, leads to the interpretation of the geometric condition (\ref{Y-classification}) in the observable regime as Heisenberg's uncertainty principle \cite{Gog, Go-Sa}.

In this paper we want to study connections of the 15-parameter conformal symmetry of the 4D Minkowski space-time with the transformations of the 8D octonionic space (\ref{sN}). In 4D the conformal symmetry is composed of the following three basic transformations:
\begin{enumerate}
\item{Poincar\'{e} transformations:
\begin{equation} \label{Poincare}
x^\nu \to a^\nu_\mu x^\mu + b^\nu~, ~~~~~~~~ (\mu, \nu = 0, 1, 2, 3)
\end{equation}
where $a^\nu_\mu$ and $b^\nu$ are the parameters of 4D rotations and translations, respectively.}
\item{Dilatations:
\begin{equation} \label{Dilatation}
x^\nu \to c x^\nu ~,
\end{equation}
where $c$ is a dimensionless constant.}
\item{Inversions:
\begin{equation} \label{Inversion}
x^\nu \to L^2\frac {x^\nu}{x^\mu x_\mu}~,
\end{equation}
where $L$ is some constant with dimension of length.}
\end{enumerate}

Note that the octonionic 8D interval (\ref{sN}) can be written in the 6D form describing a (2+4)-cone,
\begin{equation} \label{6D}
L^2\left(\varpi^2 - l^2\right) - X^\nu X_\nu  = 0~,
\end{equation}
the linear Lorentz-type rotations of which are known to generate the 4D non-linear conformal transformations (\ref{Poincare}), (\ref{Dilatation}) and (\ref{Inversion}) \cite{Dirac}. To obtain (\ref{6D}) we used the fact that zero divisors (special elements of the algebra with zero norms) can be connected with the unit signals (elementary particles \cite{Go-Sa, Gog-PP}) and we had introduced the new coordinates,
\begin{equation} \label{X}
\varpi = \frac 1L ~\omega~, ~~~~~ l = \frac 1L ~\sqrt{\lambda_1^2 + \lambda_2^2 + \lambda_3^2}~, ~~~~~ X^\nu = \frac {\sqrt 2}{2}(\varpi + l) x^\nu ~.
\end{equation}
In terms of the coordinates (\ref{X}) conformal transformations (\ref{Poincare}), (\ref{Dilatation}) and (\ref{Inversion}) can be written in the form \cite{Murai}:
\begin{enumerate}
\item{ Poincar\'{e} transformations:
\begin{eqnarray}
X^\nu &\to& a^\nu_\mu X^\mu + \frac {L}{\sqrt 2}b^\mu(\varpi +l)~, \nonumber \\
\varpi &\to& \varpi - \frac {\sqrt 2}{2 L} g_{\alpha \beta} a^\alpha_\gamma b^\beta X^\gamma - \frac {\sqrt 2}{4}  g_{\alpha \beta} b^\alpha b^\beta(\varpi + l)~, \\
l &\to& l + \frac {\sqrt 2}{2 L} g_{\alpha \beta} a^\alpha_\gamma b^\beta X^\gamma + \frac {\sqrt 2}{4} g_{\alpha \beta} b^\alpha b^\beta(\varpi + l)~. \nonumber
\end{eqnarray}
}
\item{ Dilatations:
\begin{eqnarray}
X^\nu &\to& X^\mu ~, \nonumber \\
\varpi &\to& \frac 12 \left(c + \frac 1c \right) \varpi - \frac 12 \left(c - \frac 1c \right) l~, \\
l &\to& \frac 12 \left(c + \frac 1c \right) l - \frac 12 \left(c - \frac 1c \right) \varpi ~. \nonumber
\end{eqnarray}
}
\item{ Inversions:
\begin{eqnarray}
X^\nu &\to& X^\nu ~, \nonumber \\
\varpi &\to& \varpi ~, \\
l &\to& -l ~. \nonumber
\end{eqnarray}
}
\end{enumerate}

Now let us explore the relationships between the active transformations of the octonionic 8D space (\ref{sN}) with the conformal group in 4D. In general, 8D quadratic forms like (\ref{sN}) are invariant under the group $SO(4,4)$ of tensorial transformations of their parameters. However, if we want to represent rotations by octonions, i.e. to perform active transformations of octonionic basis units, we need automorphisms \cite{Go-Sa}. The group of automorphism of split octonions is the real non-compact form of Cartan's smallest exceptional Lie group $G_2^{NC}$ \cite{Cart, BHW, Man-Sch}, which is the main geometrical group of our model \cite{Gog, Go-Sa, Gog-PP}. Infinitesimal transformations of coordinates of (\ref{s}), which correspond to the $G_2^{NC}$-rotations, can be written as \cite{Go-Sa}:
\begin{eqnarray} \label{x-nu}
x_n' &=& x_n - \varepsilon_{nmk} \alpha^m x^k - \theta_n ct + \frac 12 \left( |\varepsilon_{nmk}|\phi^m + \varepsilon_{nmk} \theta^m \right) \lambda^k + \left(\varphi_n - \frac 13 \sum_m \varphi_m\right) \lambda_n~, \nonumber \\
ct' &=& ct - \beta_n \lambda ^n - \theta_nx^n ~, ~~~~~~~~~~~~~~~~~~~~~~~~~~~~~~~~~~~~~~~~~~~~~~~ (n, m, k = 1, 2, 3)\\
\lambda_n' &=& \lambda_n - \varepsilon_{nmk} \left(\alpha^m - \beta^m\right) \lambda^k + \beta_n ct + \frac 12 \left( |\varepsilon_{nmk}| \phi^m - \varepsilon_{nmk} \theta^m \right) x^k + \left(\varphi_n - \frac 13 \sum_m \varphi_m\right) x_n~, \nonumber
\end{eqnarray}
no summing over $n$ in the last terms of $x_n'$ and $\lambda_n'$. From fifteen parameters in (\ref{x-nu}) (the five 3-angles: $\alpha^m$, $\beta^m$, $\phi^m$, $\theta^m$ and $\varphi^m$), due to the condition
\begin{equation} \label{varphi-1/3}
\sum_n \left(\varphi_n - \frac 13 \sum_m \varphi_m\right) = 0~,
\end{equation}
only 14 are independent.

We notice that the $G_2^{NC}$-rotations (\ref{x-nu}), which affect only the 7D vector part of split octonions, in 4D sub-space generate some kind of Poincar\'{e} transformations. Rotations by $\alpha^n$ and $\theta^n$ represent generalized Lorentz transformation (with the Lorentz factor (\ref{|V|})), the angles $\beta^n$ give time translations, while the hyperbolic rotations of $x^n$ by the angles $\phi^n$ and $\varphi^n$ towards the time-like extra dimensions $\lambda_n$ correspond to the spatial translations \cite{Go-Sa}. Note that 14-dimensional algebra of $G_2^{NC}$ does not reduce to the exact Lorentz algebra $SO(3,1)$, in fact $G_2^{NC}$ has not any 6D subalgebra \cite{BHW}. However, there exists the subalgebra $SU(2,1)$, with the elements having similar to $SO(3,1)$ commutators, apart from diagonal combinations of rotations and boosts, which commute in $SO(3,1)$, but not in $SU(2,1)$. Thus the six $G_2^{NC}$-Lorentz-type elements do not close, what would predict new physical effects in the model, like parity violation \cite{Go-Sa}. Another difference is that time translations are cyclic, since correspond to the compact angles $\beta^n$, while there exists some maximal spatial translations generated by the hyperbolic angles $\phi^n$ and $\varphi^n$, leading to the appearance of spatial horizons.

Now let us focus on the $G_2^{NC}$-rotations which in 4D-subspace generate dilatations and inversions. Consider mutual rotations of a pair of spatial and time-like coordinates, $x^n$ and $\lambda^n$, by the hyperbolic angles, $\varphi^n$ and $\phi^n$, which preserve the form $x^nx_n - \lambda^n\lambda_n$. These $G_2^{NC}$-transformations correspond to the different types of coordinate transformations for the cases listed in (\ref{Y-classification}):
\begin{itemize}
\item{In the observable regime the $G_2^{NC}$-rotations that preserve the 2D form $x^2 - \lambda^2$ can be represented as:
\begin{eqnarray} \label{observable}
\lambda' &\to& \frac {2x\lambda}{x^2 - \lambda^2} ~, \nonumber \\
x' &\to& \frac {x^2 + \lambda^2}{x^2 - \lambda^2} ~.
\end{eqnarray}
In 4D sub-space these transformations correspond to the spatial dilatations, $x'_n \to c x_n$, with some constant $c$.}

\item{For the virtual case, i.e. when $\lambda^2$ exceeds $x^2$, we have,
\begin{eqnarray} \label{virtual}
\lambda' &\to& \frac {x^2 + \lambda^2}{\lambda^2 - x^2} ~, \nonumber \\
x' &\to& \frac {2x\lambda}{\lambda^2 - x^2} ~.
\end{eqnarray}
Since in our model the extra time-like parameters are connected with wavelengths of particles, $\lambda^n \sim \hbar/p_n$ \cite{Gog, Go-Sa, Gog-PP}, and the true time parameter $t$ is not affected by the angles $\varphi^n$ and $\phi^n$, in 4D the transformations (\ref{virtual}) correspond to the spatial inversions $x_n' \to tx_n/x^2$.}
\end{itemize}
So linear $G_2^{NC}$-automorphisms of the octonionic 8D space in the 4D sub-space generate some analogues to the complete non-linear conformal transformations.

At the end of the paper let us estimate the physical value of the parameter $L$ (characteristic length), which we used to define dimensionless coordinates in (\ref{X}). From the inequalities (\ref{|V|}) and (\ref{Y-classification}) it follows that in the observable regime, together with the existence of the light horizon $D_H$, we have the maximal wavelength,
\begin{equation}
|\lambda^n|_{max} \leq D_H \sim 10^{26} ~m ~.
\end{equation}
As it was mentioned above, in our approach elementary particles are corresponded to the light-cone (zero-norm) signals \cite{Gog-PP}. Then $G_2^{NC}$-rotations of the (3+4)-sphere with the three extra time-like dimensions, $\lambda^n$, can be viewed as a motion of a particle in the effective de Sitter space-time,
\begin{equation} \label{deSitt}
-c^2 t^2 + x^nx_n = L^2~,
\end{equation}
where the constant
\begin{equation}
L = |\lambda^n|_{max}
\end{equation}
naturally corresponds to the observed value of the 4D cosmological constant $\Lambda$,
\begin{equation}
L^2 \sim \frac 1\Lambda \sim 10^{52} ~m~.
\end{equation}

To conclude, in this paper we have studied relationships of the conformal symmetries of the 4D Minkowski space with the transformations of the non-compact form of Cartan exceptional Lie group $G_2$, the main space-time group of the octonionic geometry model. It was shown that the dimensional constant that appears in these analysis naturally corresponds to the observed value of the cosmological constant.


\end{document}